\documentstyle[12pt,moriond,epsf]{article}
\begin{document}
%
% Definitions for this paper
\newcommand{\Halpha}{\mbox{H$\alpha$}}
\newcommand{\Oii}{\mbox{[O$_{II}$]}}
\newcommand{\ew}{\mbox{$W_\lambda$}}
\newcommand{\ewh}{\mbox{$W_\lambda$ H$\alpha$}}
\newcommand{\ewo}{\mbox{$W_\lambda$ [O$_{II}$]}}
\newcommand{\etal}{{\em et al.}}
\newcommand{\hMpc}{h^{-1}{\rm Mpc}}
\newcommand{\hcMpc}{h^{-3}{\rm Mpc^3}}
\newcommand{\hcMpcinv}{h^{3}{\rm Mpc}^{-3}}
\newcommand{\rmsub}[1]{\mbox{$_{\rm #1}$}}	% roman subscript

\heading{THE FAINT END OF THE LUMINOSITY FUNCTION IN THE FIELD}

\author{Jon Loveday} {University of Chicago, Chicago, USA}

\begin{abstract}
\baselineskip 4.5 true mm % GAM addition
I review the current observational status of the faint end of the optical 
luminosity function of field galaxies at low redshift.
There is growing evidence for an excess number of dwarf galaxies that is
not well fit by a single Schechter function.
These dwarf galaxies tend to be of late morphological and spectral type, 
blue in colour,
of low surface brightness and currently undergoing significant star formation.
\end{abstract}

\section{Introduction}
The galaxy luminosity function (LF) characterizes the number density of
galaxies as a function of luminosity, or absolute magnitude.
The faint end of the LF tells us the abundance of dwarf galaxies,
accurate knowledge of which is important for constraining 
models of both galaxy formation and evolution.

The optical galaxy luminosity function has traditionally been fit by
a Schechter \cite{s76} function:
\begin{equation}
\phi(L)dL = \phi^* \left(\frac{L}{L^*}\right)^\alpha 
            \exp\left(\frac{-L}{L^*}\right) d\left(\frac{L}{L^*}\right).
\label{eqn:schec}
\end{equation}
For galaxies more luminous than a characteristic luminosity $L^*$, $\phi(L)$
drops exponentially with luminosity; for galaxies fainter than $L^*$
$\phi(L)$ approaches a power-law with slope $\alpha$.
The quantity $\phi^*$ represents an overall normalisation of the luminosity
function.

Most local galaxy surveys (eg. \cite{eep88,lkslots96,lpem92,mhg94,rspf98})
have found a value of $\alpha$ that lies in the
range $-1.2 \lsim \alpha \lsim-0.7$, giving a ``flat'' faint-end slope 
when the LF is plotted as a function of absolute magnitude.
However, there is recent evidence that the faint end of the LF is not 
well fit by a Schechter function:
\begin{enumerate}
\item An analysis of the CfA redshift survey by Marzke \etal\ \cite{mhg94}
finds that although the best-fit Schechter function has $\alpha = -1.0 \pm 0.2$,
there is a significant excess of galaxies above the Schechter fit at magnitudes
$M_Z \gsim -16$\footnote{Throughout, absolute magnitudes will be quoted 
assuming a Hubble constant $H_0$ of 100 km s$^{-1}$ Mpc$^{-1}$}.
\item For the Las Campanas Redshift Survey, Lin \etal\ \cite{lkslots96}
measure a faint-end slope of $\alpha = -0.70 \pm 0.05$ in the Gunn-$r$ band.
However, all of their data points fainter than $M_r \approx -17.5$ lie
above this Schechter fit.
\item In their analysis of the ESO Slice Project (ESP) redshift 
survey, Zucca \etal\ \cite{z97} find that the $b_J$ luminosity function
is best fit by a flat faint-end Schechter function for $M_{b_J} < -17$
and a power-law of slope $-1.57$ at fainter magnitudes.
\end{enumerate}

Given the importance of the number-density of local dwarf galaxies for theories
of galaxy formation and evolution, it is extremely desirable to better
constrain the faint-end of the luminosity function.
The problem is that most galaxies in flux-limited surveys have $L \sim L^*$
and measuring galaxy redshifts to fainter magnitudes 
does not increase the number
of dwarf galaxies {\em relative} to the number of bright galaxies.
There is no substitute for surveying large numbers of faint galaxies to 
constrain  the faint-end of the LF.  
Thus we require either to carry out large redshift surveys, such as 
2dF~\cite{c98} and SDSS~\cite{l96}, or to estimate galaxy redshifts 
from their photometric colours or clustering properties.

In the following section I will describe how one can constrain the space
density of dwarf galaxies from counts of faint galaxies around
bright galaxies of known redshift.
I will then review the properties of galaxies which dominate
the faint-end of the LF and discuss future prospects for measuring 
the field galaxy LF.

\section{Constraining the Space Density of Dwarf Galaxies}
\label{sec:clust}

If one has a large, photometric galaxy sample, and redshifts for a
subset of these galaxies, one can make use of the fact that excess
galaxies seen close in projection to galaxies of known redshift are
likely to be at the same distance.
This is a statistical generalization of the method used to determine
the luminosity function of galaxies in clusters in which one assumes
that all galaxies in a cluster are at the same distance from the observer.
Due to the large numbers of galaxies, one can estimate $\phi(M)$ to
faint luminosities \cite{ps87,l97}.

\subsection{Method}

Consider in turn each redshift survey (centre) galaxy at distance $y_i$:
\begin{itemize}
\item Count galaxies in bins of projected separation 
$\sigma = y_i \theta$ (where $\theta$ is the angular separation)
and apparent magnitude $m$ about each centre.
\item The excess counts above a random distribution gives an estimate of the 
product
$X(\sigma, M) = \Xi(\sigma) \times \phi(M)$, where $\Xi(\sigma)$ is the 
projection of the spatial correlation function $\xi(r)$ along the line of sight:
$$\Xi(\sigma) = \int_{-\infty}^{+\infty} 
\xi( \sqrt{ \Delta y^2 + \sigma ^2 } ) d \Delta y.$$
\end{itemize}
One then averages $X(\sigma, M)$ over all centre galaxies with 
minimum-variance weighting and
divides $X(\sigma, M)$ by an {\em assumed} $\Xi(\sigma)$ 
to obtain a final estimate of $\phi(M)$.

\begin{figure}[htbp]
\begin{center}
\leavevmode
\epsfverbosetrue
\epsfxsize=0.6\textwidth
\epsfbox{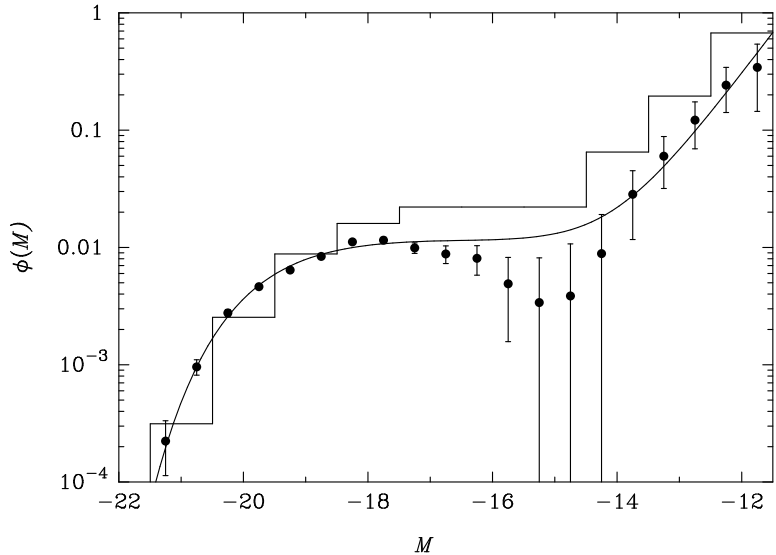}
\end{center}
\caption[]{Estimate of the galaxy luminosity function made by counting
faint APM galaxies around Stromlo-APM redshift survey galaxies.
The smooth curve shows the best-fit ``double power-law'' Schechter function
(\ref{eqn:dpschec}) and the histogram shows the Gronwall \& Koo 
model \cite{gk95}.}
\label{fig:lumx}
\end{figure}

\subsection{Results}

The above analysis was carried out using 2.4 million APM galaxies and 1787
centre galaxies from the Stromlo-APM redshift survey \cite{l97}.
The estimated LF is shown in Figure~\ref{fig:lumx}, where we have counted
faint galaxies to a projected separation $\sigma \le 1\hMpc$ and we have 
assumed that the galaxy correlation function is given by the power-law
$\xi(r) = (r/5.1\hMpc)^{-1.71}$.
We see two major differences from a flat faint-end Schechter function:
\begin{enumerate}
\item The estimated LF drops below a flat faint-end Schechter function in
the luminosity range $-17 \le M_{b_J} \le -14$.  
This is most likely due to the fact that this analysis assumes galaxies 
cluster independently of their luminosity, whereas we know that
sub-$L^*$ galaxies are more weakly clustered than $\sim L^*$ galaxies
\cite{lmep95}.
\item Faintward of $M \approx -15$, the LF steepens and exceeds a
flat faint-end Schechter function by $M = -14$.
\end{enumerate}
To model the observed $\phi(M)$, we have fitted a modified form of
the Schechter function, with an additional faint-end power law:
\begin{equation}
\phi(L) = \phi^* \left(\frac{L}{L^*}\right)^\alpha 
	    \exp\left(\frac{-L}{L^*}\right)
	    \left[1 + \left(\frac{L}{L_t}\right)^\beta\right].
\label{eqn:dpschec}
\end{equation}
In this formulation $\phi^*$, $L^*$ and $\alpha$ are the standard Schechter
parameters, $L_t$ is a transition luminosity between the two power-laws
and $(\alpha + \beta)$ is the power-law slope of the very faint-end.
No physical interpretation is intended by this choice of formula,
it is merely a convenient way of modeling the observed $\phi(L)$
over this extended range of luminosity and for estimating the faint-end
slope.
The line in Figure~\ref{fig:lumx} shows the best-fit
``double power-law'' luminosity function, which has parameters:
$\alpha = -0.94$, $M^* = -19.65$, $\phi^* = 0.0154 \hcMpc$, 
$M_t = -14.07$ and $\beta = -1.82$ (not $\beta = -2.82$ as erroneously
stated in \cite{l97}).
Although this fit is poor over the range $-17 \lsim M \lsim -14$,
our $\phi(M)$ estimate is almost certainly biased low
over this range by the weaker clustering of galaxies fainter than $L^*$.
Clearly, the faint-end slope $\alpha + \beta \approx -2.8$ cannot extend to 
indefinitely
low luminosities, but it shows no obvious signs of flattening brightward
of $M = -12$.

Since we are essentially measuring the product of the galaxy luminosity 
function $\phi(M)$ with the projected correlation function $\Xi(\sigma)$,
then any dependence of $\Xi(\sigma)$ on galaxy luminosity can bias our
estimate of $\phi(M)$.
Thus the faint-end turn up could be due to a genuine increase in the
local density of dwarf galaxies or an artifact caused by strong clustering
of dwarf galaxies around $L^*$ galaxies.

However, one can place a lower limit on the space density of dwarf
galaxies by making the extreme assumption that they only exist close
to $\sim L^*$ galaxies.  We see an excess of dwarf galaxies up to a
projected separation of at least $5 \hMpc$ from $\sim L^*$ galaxies
\cite{l97}.  Thus a limit may be estimated by assuming that they
occur with the measured space density only within $5 \hMpc$ of an
$\sim L^*$ galaxy.  Integrating (\ref{eqn:dpschec}) over the magnitude range
$-15 \le M_{b_J} \le -12$, we measure a space density 
$\bar{n} \approx 0.20 \hcMpcinv$.
A lower limit on the average mean density of
dwarf galaxies is then given by multiplying this
by the fraction of space within $5 \hMpc$ of an $\sim L^*$
galaxy ($\approx 0.6$).  We thus arrive at a limit on the space density
of dwarf galaxies, here defined to lie in the luminosity range 
$-15 \le M_{b_J} \le -12$, of $\bar{n} \gsim 0.12 \hcMpcinv$.  
This is a factor of two higher
than the density $\bar{n} \approx 0.058 \hcMpcinv$ inferred from the
extrapolation of a
$\alpha = -1.11$ Schechter function \cite{lpem92}.

In fact, the true space density of dwarf galaxies is likely to be
significantly higher than this, for a number of reasons.
First, our estimator assumes that galaxy clustering is independent of 
luminosity, whereas we know that sub-$L^*$ ($-19 < M < -15$) 
galaxies are less strongly clustered
than more luminous galaxies \cite{lmep95}.
If this luminosity segregation extends to dwarf ($M > -15$) galaxies, 
then we will have underestimated their space density.
Second, analysis of simulations \cite{l97} shows that our estimator
tends to underestimate the faint-end slope of $\phi(M)$ slightly,
possibly due to a Malmquist-type bias.
Third, as discussed by numerous authors (eg. \cite{siib97} and references
therein), most galaxy surveys
are likely to be missing a substantial fraction of low surface brightness
galaxies, many of which will be dwarfs.
For example, Sprayberry \etal\ \cite{siib97} find a pronounced upturn in the
luminosity function for their sample of low surface-brightness galaxies.
Thus the true space density of dwarf galaxies in the magnitude range
$-15 \le M_{b_J} \le -12$ could be significantly higher than
$\bar{n} = 0.12 \hcMpcinv$.

A high space density of dwarf galaxies, assuming that they are predominantly
late-type, blue galaxies (see following section), which suffer smaller 
$K$-correction dimming
than redder, early-type galaxies, helps provide a natural explanation for
the steep observed number counts of faint galaxies.
Gronwall and Koo \cite{gk95} were able to match observations of galaxy 
number counts
in the $K$, $R$ and $B_J$ bands, as well as colour and redshift distributions,
for a mild evolution model by assuming that the faint end of the galaxy 
luminosity function is dominated by blue galaxies ($B - V \le 0.6$),
and rises significantly above a Schechter function with flat faint-end slope.
The Gronwall \& Koo model total luminosity function is plotted in
Figure~\ref{fig:lumx}, and one sees remarkably good agreement with
our observations.
These results thus support the argument of Gronwall \& Koo; 
there is no need to invoke
exotic forms of galaxy evolution to explain observed galaxy number counts
at faint magnitudes.

As a final caveat, I would stress that all existing data on the LF of galaxies
fainter than $M \approx -15$ comes from galaxies within a distance of 
$115\hMpc$ (within $30\hMpc$ for $M \ge -12$)
and so the estimated faint-end LF is susceptible to sampling fluctuations.
More data is certainly needed before we can really claim to have nailed down
the faint end of the galaxy luminosity function.

\section{Properties of Dwarf Galaxies in the Field}

\begin{figure}[htbp]
\begin{center}
\leavevmode
\epsfxsize=0.6\textwidth
\epsfbox{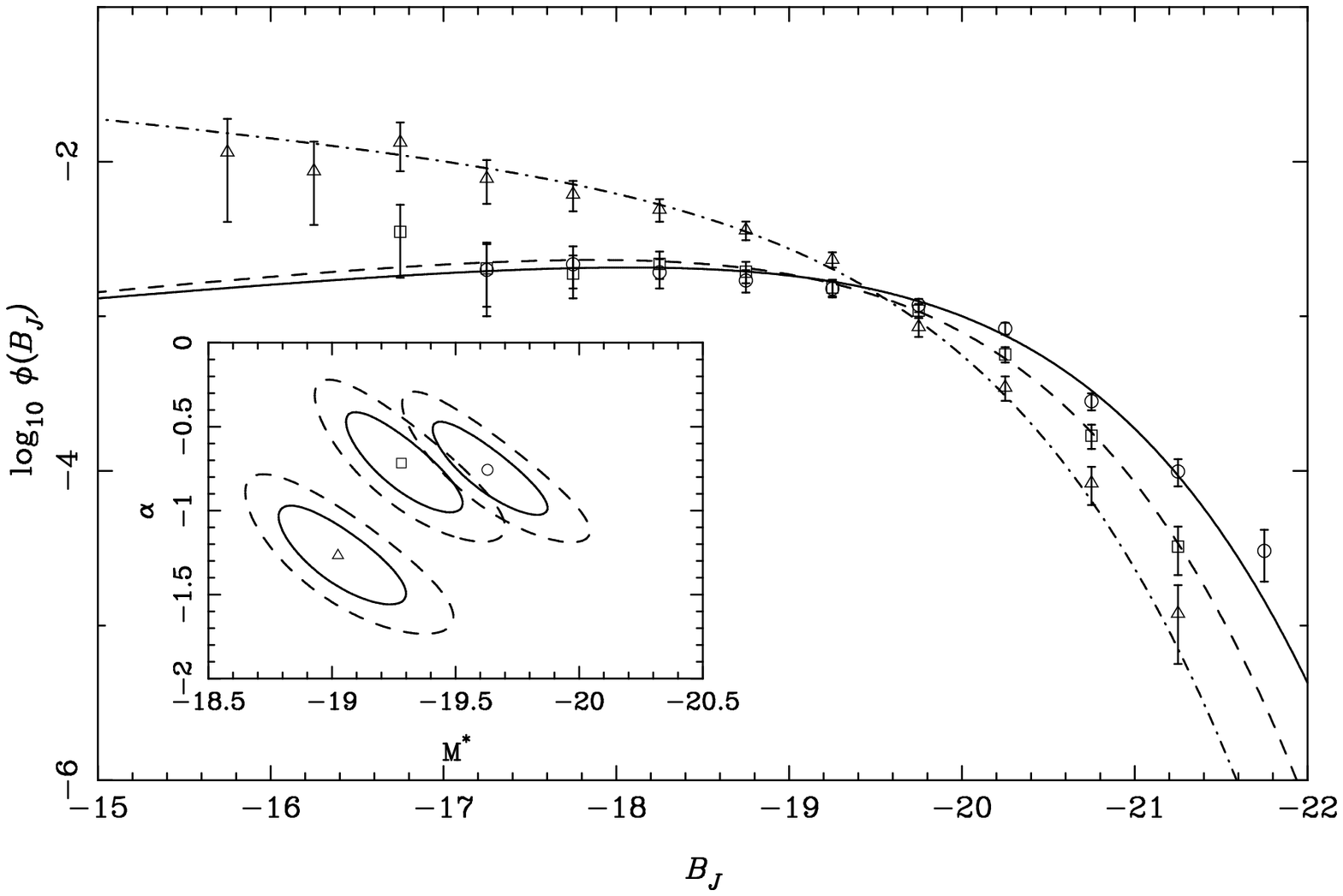}
\end{center}
\caption[]{Estimates of the luminosity function for galaxies in the
Stromlo-APM survey with no
significant detected \Halpha\ emission ($\ewh < 2$\AA: circles, solid line),
with moderate \Halpha\ emission (2\AA\ $\le \ewh < 15$\AA: squares, dashed
line) and with strong \Halpha\ emission ($\ewh \ge 15$\AA: 
triangles, dot-dashed line).
For clarity, data points representing fewer than five galaxies have been
omitted from the plot.
The inset shows 1 \& 2$\sigma$ likelihood contours for the best-fit 
Schechter parameters.}
\label{fig:lf_ha}
\end{figure}

In this section I review measurements of the luminosity function of
galaxy samples selected by morphological and spectral type, colour, 
surface brightness
and star-formation activity to investigate the properties of galaxies
that dominate the faint-end of the LF.
The faint-end slope $\alpha$ in what follows refers to a standard
Schechter function (\ref{eqn:schec}) fit performed by the respective authors.
\begin{description}
\item[Morphology]
Marzke \etal\ \cite{mghc94} have measured the luminosity function
of galaxies in the CfA redshift surveys selected by morphological type.
They find that the LF of galaxies with Hubble types ranging from
E to Sd all have a flat faint-end slope, but that the LF of Magellanic 
spirals and irregulars has faint-end slope $\alpha = -1.87$.
Very similar results are found from a recent analysis of the Southern Sky 
Redshift Survey (SSRS, \cite{mcpwg98}), in which the LF of irregular/peculiar
galaxies has faint-end slope $\alpha = -1.81 \pm 0.24$.
\item[Spectral Type]
Bromley \etal\ \cite{bplk98} have classified galaxies in the Las Campanas
Redshift Survey by spectral type.  The faint-end slope $\alpha$
is seen to steepen systematically and significantly from earliest
($0.54 \pm 0.14$) to latest ($-1.84 \pm 0.11$) spectral classification.
These authors also find that the faint-end slope steepens with local
density for early type-galaxies; for late-type galaxies there is no
significant luminosity-density relation.
A steepening of faint-end slope with later spectral types is also
seen in a preliminary analysis of data from the 2dF Redshift Survey \cite{c98}.
\item[Colour]
Marzke \& da Costa \cite{mc97} have subdivided the 
SSRS into blue and red samples at $(B-R) = 1.3$.
The blue sample is well-fitted by a Schechter function with faint-end
slope $\alpha = -1.51 \pm 0.18$; the red sample with $\alpha = -0.67 \pm 0.24$.
\item[Surface Brightness]
Sprayberry \etal\ \cite{siib97} have measured the luminosity function
from an APM survey of low surface brightness (LSB) galaxies.
Fitting a Schechter function to their LF they measure $\alpha = -1.42$.
However, their data is better fit by a Schechter function with $\alpha = -0.92$
for galaxies brighter than $M_B = -16$ and a power law with slope $-2.20$ for
fainter galaxies.
They find that the space density of LSBs exceeds that of ``normal'' high
surface-brightness galaxies for $M_B \ge -15$.
\item[Star Formation Activity]
Several authors have recently measured the LF of galaxies selected by
star-formation activity.
Lin \etal\ \cite{lkslots96} subdivided the Las Campanas Redshift Survey
into two using the equivalent width of the $[O_{II}] 3727$ line
and found that star-forming galaxies ($W_\lambda > 5$\AA) have 
$\alpha = -0.9 \pm 0.1$ versus $\alpha = -0.3 \pm 0.1$ for non star-forming 
galaxies.
The later spectral types of \cite{bplk98} also correspond to strong 
emission-line galaxies.
Similarly, Zucca \etal\ \cite{z97} find a significantly steeper faint-end
slope for galaxies with detected $[O_{II}]$ in the ESP survey.
We \cite{ltm98} have recently measured the LF of galaxies in the Stromlo-APM 
survey selected by equivalent width of both \Halpha\ and \Oii\ lines.
The LFs for the \Halpha\ selected samples are shown in Figure~\ref{fig:lf_ha};
we see very similar results for the \Oii\ selected samples.
We find that galaxies with significant star-formation dominate the luminosity
function fainter than $M_{b_J} \approx -19$.
\end{description}

It is thus apparent that the faint-end of the field galaxy 
luminosity function is dominated by galaxies which tend to be of
late morphological and spectral type, blue in colour, of low surface-brightness
and which are currently undergoing significant star-formation.
These results strongly suggest that it is dwarf irregular galaxies, rather
than dwarf ellipticals, that dominate the faint end of the luminosity function
in the field.
This contrasts with dwarf galaxies in cluster environments, which are mostly
dwarf ellipticals (Trentham, these proceedings).

\section{Future Prospects}

\begin{figure}[htbp]
\begin{center}
\leavevmode
\epsfxsize=0.6\textwidth
\epsfbox{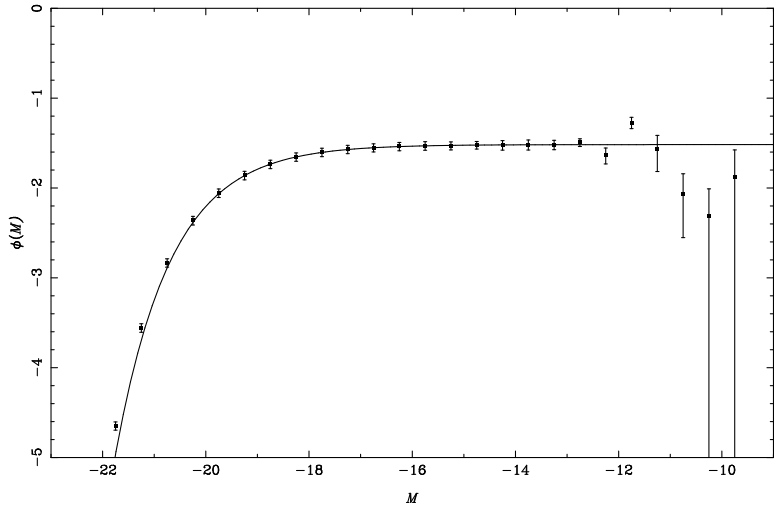}
\end{center}
\caption[]{Simulated LF we expect to measure from the Sloan Digital Sky Survey
spectroscopic galaxy sample assuming that the true LF is given by a Schechter
function with $\alpha = -1.0$, $M^* = -19.5$ (smooth curve).
Note that we will be able to reliably measure the LF to 7.5 magnitudes
fainter than $M^*$.}
\label{fig:sdss}
\end{figure}

In the near future, two large new redshifts surveys will have a dramatic
impact on measurement of the field LF.
The Anglo-Australian Telescope 2dF Galaxy Redshift Survey \cite{c98}
will measure redshifts for $\sim 250,000$ galaxies to $b_J = 19.5$
and for 100,000 galaxies to $R = 21$.
The Sloan Digital Sky Survey (SDSS \cite{l96}) will image $\pi$ sr of the
northern sky in 5 colours to $r^\prime \approx 23$ and measure redshifts 
for $10^6$ galaxies.
I have made five simulations of the SDSS main galaxy spectroscopic sample
(900,000 galaxies to $r^\prime = 18$) assuming a Schechter LF with
$\alpha = -1.0$ and $M^* = -19.5$.
The LF for each simulation was estimated using the stepwise maximum likelihood
estimator \cite{eep88} and in Figure~\ref{fig:sdss} I show the mean estimated
LF along with the rms dispersion between the simulations.
One can see that we can expect to measure a reliable LF to
$M_{lim} \approx M^* + 7.5$, or even fainter if the LF does indeed have
a steeper faint-end slope.

One would expect to be able to determine the LF to $\approx M^* + 11$
by measuring the clustering of faint galaxies in the SDSS about galaxies 
with measured redshifts.
The large SDSS dataset and accurate photometry will enable
the clustering of galaxies as a function of luminosity as well as separation
$\xi(r, L_1, L_2)$ to be accurately determined.
Thus the uncertainty
in the luminosity-dependence of the correlation function,
which dominates the errors in constraining the
space density of dwarf galaxies described in \S\ref{sec:clust}, will be
avoided.

Photometric redshifts estimated from the five SDSS colours
(eg. \cite{scsk96}) will be extremely important for studying
evolution of the LF, but of limited use for the local faint-end since 
the uncertainty in photometric redshift $\delta z\rmsub{phot} \approx 0.03$.
Possibly the best determination of the faint-end of the field galaxy LF 
will come from spectroscopic followup of faint $z\rmsub{phot} \lsim 0.1$ 
galaxies selected from the SDSS.

As emphasized elsewhere \cite{bst88,bplk98}, galaxies of different 
morphological or spectral type have different luminosity functions,
and the LF for each galaxy type should be measured independently.
The Sloan Digital Sky Survey will provide a wealth of morphological,
colour and spectral information for each galaxy and thus enable a 
careful study of the dependence of the LF on galaxy properties.

The simulated LF in Figure~\ref{fig:sdss} assumes that the SDSS
galaxy sample will be purely flux-limited.
In practice of course, surface brightness also plays an important role
in galaxy detection.
For example, Sprayberry \etal\ \cite{siib97} argue that most existing
local galaxy catalogues have missed a substantial fraction of (LSB)
galaxies.
Since the SDSS imaging will be carried out in drift-scan mode, detection
of LSB objects will be limited by photon statistics rather than flat-fielding
errors.
We expect to be able to detect at $5\sigma$ a $r^\prime = 19.5$ galaxy 
with a scale-length of $16^{\prime\prime}$ and a central surface brightness 
of 27.5 mag arcsec$^{-2}$ \cite{bb97}.
Even though we will not be able (or even attempt) to measure redshifts 
for most galaxies
of this surface brightness, the surface brightness selection criteria
will be very well defined and therefore correctable.

\section{Conclusions}

\begin{enumerate}
\item Evidence is building that a single Schechter function cannot fit the 
observed LF over a wide range of magnitudes.
This is at least partly due to the fact that galaxies of different 
morphological type have differing LFs.
\item Several surveys are finding excess galaxies above a 
Schechter function at the faint end.
\item These excess galaxies tend to be of late morphological and spectral type,
blue in colour, of low surface-brightness and undergoing significant 
star-formation.
The faint-end of the field LF is thus dominated by dwarf irregular rather than
dwarf elliptical galaxies.
\item A turnup in the LF of dwarf galaxies can help reconcile faint galaxy 
number counts and redshift distributions without the need to invoke
exotic evolutionary models.
\item Future surveys, such as The Sloan Digital Sky Survey (SDSS),
will enable us to measure the field galaxy LF to $M \approx M^* + 11$,
provided proper account is taken of surface brightness selection effects.
\end{enumerate}

It is a pleasure to thank the organizers for a most enjoyable meeting and for
providing financial support.

\end{document}